\documentclass[letter,traditabstract]{aa}
\usepackage{graphicx}
\usepackage{txfonts}
\usepackage{todonotes}
\usepackage[normalem]{ulem}
\usepackage{amstext}
\usepackage[breaklinks=true]{hyperref}
\usepackage{xspace}
\usepackage{lscape}
\usepackage{mathrsfs}
\usepackage{color}
\usepackage{placeins}

\usepackage{listings}
\makeatletter
\renewcommand*\maketitle{%
  \thispagestyle{firstpage}
\begingroup
    \if@wideboxfn
    \setlength\bibindent{1.4\parindent}
    \else
    \setlength\bibindent{\parindent}
    \fi
    \renewcommand*\thefootnote{\@fnsymbol\c@footnote}%
    \renewcommand\@makefntext[1]{%
    \ifaa@longfn\hsize\textwidth\fi
    \noindent
    \hb@xt@\bibindent{\hss\@makefnmark\enspace}##1}
  \ifaa@twocolumn
  \begingroup
    \begin{aa@strip}
          \aa@maketitle
    \end{aa@strip}
    \@thanks            
  \endgroup
  \else
    \begingroup
      \let\thanks\footnote
      \aa@maketitle
    \endgroup
  \fi
\endgroup
  \setcounter{footnote}{0}%
}
\makeatother

\definecolor{dkgreen}{rgb}{0,0.6,0}
\definecolor{gray}{rgb}{0.5,0.5,0.5}
\definecolor{mauve}{rgb}{0.58,0,0.82}
\definecolor{red}{rgb}{1,0,0}
\newcommand{\orcit}[1]{\protect\href{https://orcid.org/#1}{\protect\includegraphics[width=8pt]{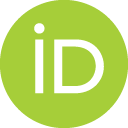}}}

\def\ltsima{$\, \buildrel < \over \sim \,$}
\def\simlt{\lower.5ex\hbox{\ltsima}}
\def\gtsima{$\, \buildrel > \over \sim \,$}
\def\simgt{\lower.5ex\hbox{\gtsima}}

\begin{document}

\title{Old massive clusters (and a nuclear star cluster?) in the tidal tails of NGC5238
} 
\subtitle{}
\authorrunning{M. Bellazzini et al.}
\titlerunning{
Old massive clusters in the tidal tails of NGC5238}

\author{
M.~Bellazzini\orcit{0000-0001-8200-810X}\inst{1}
\and
F.~Annibali\orcit{0000-0003-3758-4516}\inst{1}
\and
M. Correnti\orcit{0000-0001-6464-3257}\inst{2,6}
\and
M.~Gatto\orcit{0000-0003-4636-6457}\inst{3}
\and
M.~Marinelli\orcit{0000-0002-4775-7292}\inst{4}
\and
R.~Pascale\orcit{0000-0002-6389-6268}\inst{1}
\and
E.~Sacchi\orcit{0000-0001-5618-0109}\inst{5}
\and
M.~Tosi\orcit{0000-0002-0986-4759}\inst{1}
\and
M.~Cignoni\orcit{0000-0001-6291-6813}\inst{7,8}
\and
J.M.~Cannon\orcit{0000-0002-1821-7019}\inst{9}
\and
L.~Schisgal\orcit{0009-0007-3163-3678}\inst{9}
\and
G.~Bortolini\orcit{0009-0003-6182-8928}\inst{10}
\and 
A.~Aloisi\orcit{0000-0003-4137-882X}\inst{11,4}
\and
G.~Beccari\orcit{0000-0002-3865-9906}\inst{12}
\and
C.~Nipoti\orcit{}\inst{13}
}
\institute{
INAF - Osservatorio di Astrofisica e Scienza dello Spazio di Bologna, via Piero Gobetti 93/3, 40129 Bologna, Italy
\and
INAF - Osservatorio Astronomico di Roma, Via Frascati 33, 00078, Monteporzio Catone, Rome, Italy
\and
INAF - Osservatorio Astronomico di Capodimonte, Via Moiariello, 16, 80131 Napoli, Italy
\and
Space Telescope Science Institute 3700 San Martin Drive Baltimore, MD 21218, USA
\and
Leibniz-Institut f\"ur Astrophysik Potsdam (AIP), An der Sternwarte 16, 14482 Potsdam, Germany
\and
ASI-Space Science Data Center, Via del Politecnico, I-00133, Rome, Italy
\and
Department of Physics – University of Pisa, Largo B. Pontecorvo 3, 56127 Pisa, Italy
\and
INFN, Largo B. Pontecorvo 3, 56127, Pisa, Italy
\and
Macalester College, 1600 Grand Avenue, Saint Paul, MN 55105, USA
\and
Department of Astronomy, Stockholm University, AlbaNova University Center
\and
Astrophysics Division, Science Mission Directorate, NASA Headquarters, 300 E Street SW, Washington, DC 20546, USA
\and
European Southern Observatory, Karl-Schwarzschild-Strasse 2, 85748 Garching bei M\"unchen, Germany
\and
Dipartimento di Fisica e Astronomia “Augusto Righi”, Università di Bologna, Via Gobetti 93/2, 40129 Bologna, Italy 
}




\date{Accepted for publication on September 15, 2024}

\abstract{
New, deep HST photometry allowed us to identify and study eight compact and bright ($M_V\le -5.8$) star clusters in the outskirts of the star-forming isolated dwarf galaxy NGC~5238 ($M_{*}\simeq 10^8~M_{\sun}$). Five of these clusters are new discoveries, and six appear projected onto, and/or aligned with the tidal tails recently discovered around this galaxy. The clusters are partially resolved into stars and their colour magnitude diagrams reveal a well developed red giant branch, implying ages older than 1-2~Gyr. 
Their integrated luminosity and structural parameters are typical of classical globular clusters and one of them has $M_V=-10.56\pm 0.07$, as bright as $\omega$~Cen, the brightest globular cluster of the Milky Way. Since the properties of this cluster are in the range spanned by those of nuclear star clusters we suggest that it may be the nuclear remnant of the disrupted satellite of NGC~5238 that produced the observed tidal tails.
}


\maketitle

\section{Introduction}
\label{sec:intro}

The current $\Lambda$ - Cold Dark Matter paradigm predicts that galaxies form by means of a hierarchical merging process \citep{primack24} irrespective of the mass scale of the galaxies over a very wide range \citep{diemand05,wang20}. Hence, the paradigm must be tested at any scale, including the smallest ones, that is in the realm of dwarf galaxies. Here the test is especially challenging, due to the intrinsic faint and low surface brightness nature of the dwarfs and of their possible satellites. However, systematic searches among local unresolved galaxies begin to provide interesting results
\citep{Paudel2018,kadofong2020}.

In this context, the Smallest Scale of the Hierarchy survey \citep[SSH;][]{ssh_pap1} is aimed at searching for signs of recent interactions or merging with satellites in a sample of nearby (D$\la 10$~Mpc) isolated and (at least partially) resolved dwarf galaxies by tracing their stellar structure by star counts down to a very low surface brightness level ($\mu_r\la 31~{\rm mag/arcsec}^2)$. While requiring that the galaxies can be resolved into stars strongly limits the accessible volume, it allows to discriminate between young stars, that trace the (typically) strongly asymmetric distribution of star forming regions, and old stars, specifically Red Giant Branch (RBG) stars (age$\ga 1-2$~Gyr), whose distribution should reliably trace the galaxy gravitational potential and the possible disturbances induced by interactions with satellites.

\begin{figure*}[ht!]
\center{
\includegraphics[width=0.7\textwidth]{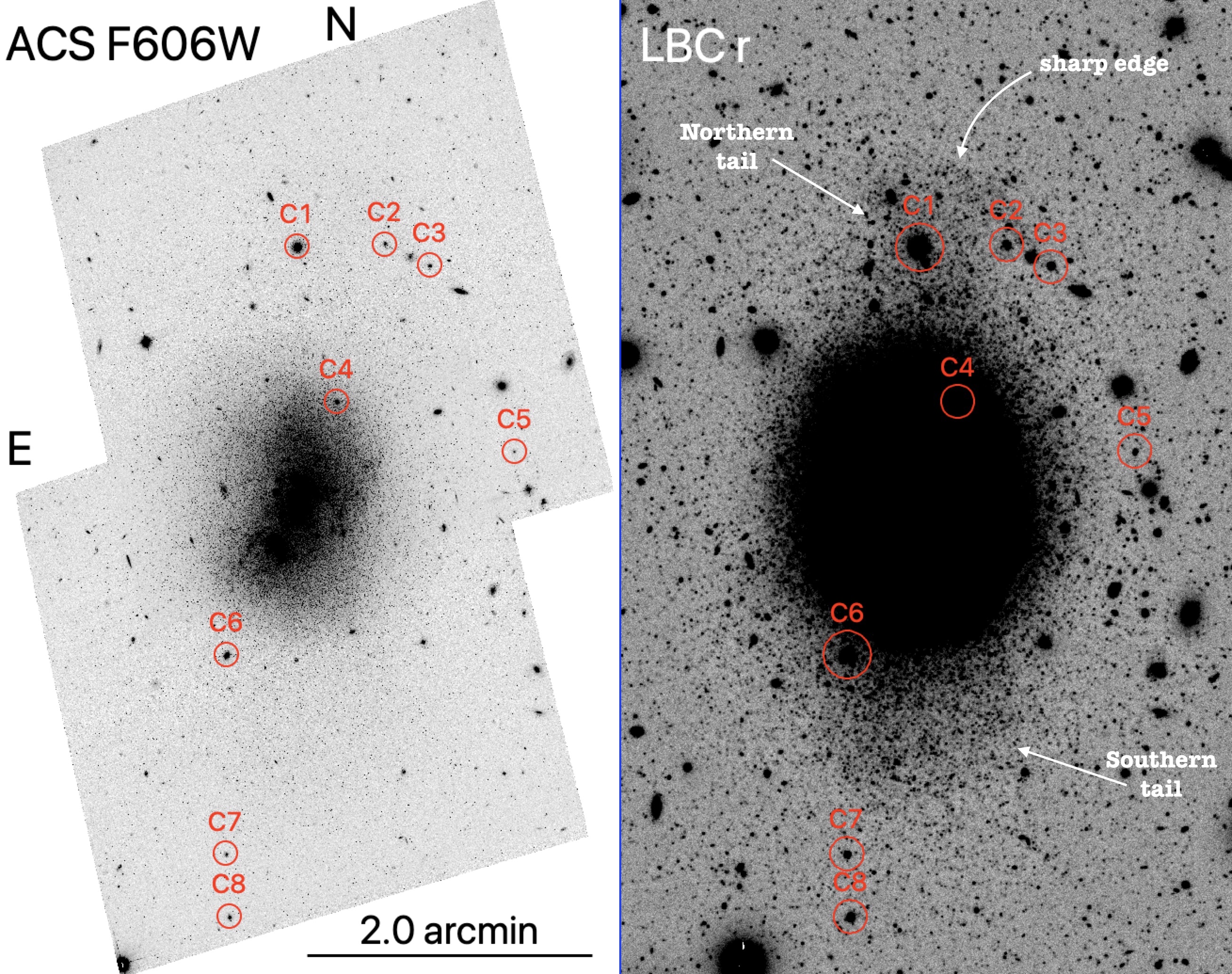}
}
\caption{Images of NGC~5238 from our HST (left image) and LBT (right image) data, with the star clusters considered in this letter circled and labelled. The different intensity cuts in the two images are intended to highlight the irregular structure of the inner star forming region (ACS image) and the regular ellipsoid of the main body as well as the shape of the tidal features protruding from its northern and southern edges (LBC image).}
\label{fig:ACSLBC}
\end{figure*}

Recently, \citet{sacchi24} reported on the discovery of prominent asymmetric extended features in the distribution of RGB stars in the outskirts of six SSH dwarfs, all interpreted as relics of the interaction with and ingestion of a former satellite \citep[e.g., similarly to the case of DDO68, see][]{annibali16,annibali19}. Among these, the star forming galaxy NGC~5238 
\citep[D$\simeq 4.5$~Mpc, $M_{*}\simeq 10^{8}~M_{\sun}$;][see Appendix~\ref{app:N5238}]{cannon16} was shown to display two wide asymmetric stellar features protruding from the northern and the southern edges of the regular ellipsoidal main body of the galaxy as traced by RGB stars (see Fig.~\ref{fig:ACSLBC}, right panel). The northern feature ends with a kind of shell with a sharp edge in surface density, somehow bending to the south-west. 
To follow up this intriguing case of ingestion of a satellite by a dwarf we obtained deep Hubble Space Telescope (HST) observations covering the entire body of the galaxy as well as the tidal tails (program GO-17140\footnote{The images of NGC~5238 from this programme have been used for a beautiful ESA image of the week; \url{https://esahubble.org/images/potw2429a/}}; P.I. F. Annibali). Inspecting the mosaic of Advanced Camera for Surveys - Wide Field Channel (ACS-WFC) images we noted, in the outskirts of the galaxy, eight obvious bright, compact and partially resolved star clusters, five of which were previously unknown. Their Colour Magnitude Diagram (CMD) suggests an old age for all of them. The newly discovered clusters, and the outermost one among those already known, lie far beyond the main body of the galaxy and appear to correlate with the tidal tails. One of them lies straight within the northern tail, at least in projection, and is as bright as the brightest globular cluster (GC) of the Milky Way (MW), showcasing properties typical of Nuclear Star Clusters \citep[NSC;][N20 hereafter]{neumayer20}.
In this letter we report on these newly discovered clusters and their relation with the tidal tails of NGC~5238.

\section{Analysis}
\label{sec:analysis}

The star cluster population in the main body of NGC~5238 has been systematically studied by 
the LEGUS HST Treasury Program (PI D. Calzetti) with an automatic pipeline \citep{legus_clus}. 
Ages were estimated by fitting Spectral Energy Distribution (SED) models to the set of available HST magnitudes, that vary from cluster to cluster. 
The vast majority of the LEGUS candidate clusters unfortunately were classified as low-quality candidates, and only nine were confirmed by visual inspection as likely genuine clusters \citep{legus_clus,cook19,cook23}. All of these nine have estimated ages $\leq 2\times10^8$ yr, according to the LEGUS catalogue\footnote{We used the reference LEGUS catalogue, that is the one that uses Milky Way extinction, averaged aperture correction method, and Padova stellar evolutionary tracks (see  \url{https://archive.stsci.edu/prepds/legus/cluster_catalogs/ngc5238.html}).}
and, above all, are deeply embedded into the main body of the galaxy.
We will re-consider these inner clusters at the light of our new data in a future contribution where we will study in detail the star formation and the morphology of the galaxy.

\begin{table*}[!htbp]
\centering
\caption{\label{tab:tab1} Compact clusters in the outskirts of NGC5238.}
{
    \begin{tabular}{cccccc}
SSH 	&   ra        &   dec	   & F606W	   & F814W & M$_V$	 \\ 
     id   	&  [deg]      &  [deg]     &  [mag]	   &	[mag]	&	[mag]  \\
\hline 
  C1            & 203.67842 & 51.64420 & 17.34$\pm$0.02  & 16.60$\pm$0.02  & -10.56$\pm$0.05\\  
  C2            & 203.66194 & 51.64461 & 20.98$\pm$0.09  & 20.17$\pm$0.09  &  -6.89$\pm$0.12\\  
  C3            & 203.65347 & 51.64202 & 20.04$\pm$0.06  & 19.34$\pm$0.06  &  -7.87$\pm$0.08\\  
  C4 (11)	& 203.67096 & 51.62609 & 19.39$\pm$0.04  & 18.65$\pm$0.05  &  -8.52$\pm$0.07\\  
  C5            & 203.63766 & 51.62032 & 22.05$\pm$0.15  & 21.25$\pm$0.15  &  -5.83$\pm$0.18\\  
  C6 (591)	& 203.69168 & 51.59655 & 18.35$\pm$0.03  & 17.61$\pm$0.03  &  -9.55$\pm$0.06\\  
  C7 (617)  	& 203.69176 & 51.57332 & 21.26$\pm$0.10  & 20.50$\pm$0.11  &  -6.63$\pm$0.13\\  
  C8            & 203.69113 & 51.56597 & 19.74$\pm$0.05  & 19.02$\pm$0.06  &  -8.17$\pm$0.06\\  
\hline
    \end{tabular}
}
\tablefoot{Numbers in parentheses in the first column are LEGUS id numbers, when available. Coordinates are J2000 as embedded in the drizzled F606W image. Magnitudes have been 
computed over circular apertures of radius $r=6.5\arcsec$ for C1, $r=2.0\arcsec$ for C4 and C5, 
and $r=3.0\arcsec$ for all the remaining clusters. 
}
\end{table*}

\begin{figure*}[ht!]
\center{
\includegraphics[width=0.8\textwidth]{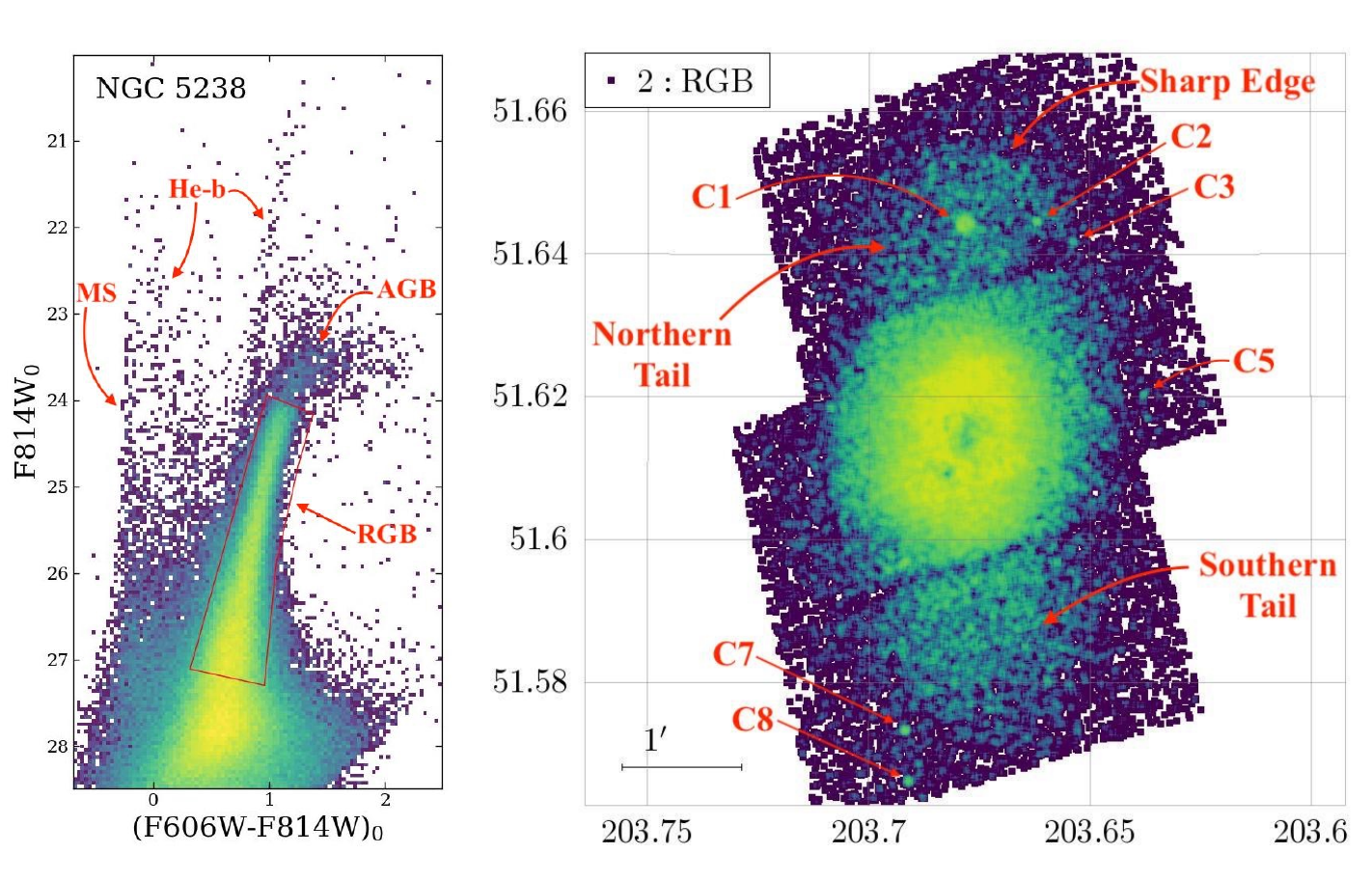}
}
\caption{Left panel: CMD of NGC~5238 from our HST ACS-WFC photometry (entire sample). The main evolutionary phases are labelled and the RGB selection box is plotted in red. Right panel: map of the RGB stars in our field (RA and Dec in degrees). Each star is colour coded according to the local density. The main tidal features are labelled as well as all the star clusters possibly associated with them, that appear as local overdensities in the map. Please note how prominent is the overdensity associated to C1. The two diagonal under-dense stripes correspond to the gaps between the ACS-WFC CCDs in the two WFC images used to build our mosaic image. In these stripes the effective exposure time is lower than elsewhere.}
\label{fig:cmdmap}
\end{figure*}

The final mosaics of drizzled ACS-WFC images of NGC~5238 we inspected summed a total exposure time of 13824~s in F606W and 22698.5~s in F814W. The F606W image is reproduced in the left panel of Fig.~\ref{fig:ACSLBC}, flanked, on the right panel, by the SSH r band image from \citet{sacchi24} obtained with the Large Binocular Camera \citep[LBC;][]{lbc} mounted on the Large Binocular Telescope (LBT\footnote{\url{https://www.lbto.org}}), exactly on the same scale.
The LBC image highlights the regular shape of the ellipsoidal main body of the galaxy and the two tails protruding from its northern and southern edges, while the intensity cuts of the ACS image have been set to reveal the complex morphology of the innermost star-forming regions. In both images we have encircled and labelled the bright and compact star clusters that we clearly identified on our ACS image as partially resolved into stars. Three of them, C4, C6 and C7, were already included in the LEGUS catalogue of candidate star clusters in NGC~5238\footnote{C6 = LEGUS 591 is among the nine confirmed LEGUS candidates, while the other two clusters in common with our sample did not pass the cut for further visual inspection in the LEGUS pipeline.}, while the other five are new discoveries, as they all lie beyond the footprint of the LEGUS images of this galaxy. Stamp images zoomed on the clusters can be found in Appendix~\ref{app:cluima}.

We used the Aperture Photometry Tool \citep[APT;][]{apt1,apt2} to estimate the cluster centres and to get surface aperture photometry on circular apertures for all the eight clusters. Throughout the paper we always report F606W and F814W magnitudes in the VEGAMAG system \citep{sirianni,bedin05}. The integrated magnitude in F606W and F814W was obtained using apertures large enough to include also the contribution of the resolved stars in the cluster outskirts (see Tab.~\ref{tab:tab1}). The background flux to be subtracted was estimated in large concentric annuli to include the contribution of resolved stars in the field of NGC~5238 as well as of background sources (the field is rich of distant galaxies). The coordinates of the clusters and their integrated magnitudes obtained in this way are reported in Table~\ref{tab:tab1}. The magnitudes of C4 and C6 are in good agreement with those measured by the LEGUS pipeline \citep[based on Sextractor;][see App.~\ref{app:sbprof}]{sextractor}, while for C7 they provide values fainter than ours by about two magnitudes, in both passbands, clearly not compatible with the observed cluster. When reducing our ACS images with Sextractor we noted that this relatively sparse and resolved cluster was split by the code into (at least) two fainter sources. This behaviour is likely at the origin of the underestimate of the integrated flux of C7 by LEGUS.

The clusters' surface brightness (SB) profiles were obtained by aperture photometry on circular annuli with APT.
We determined the structural parameters of the clusters in two ways, on the F606W profiles. First, we fitted the observed surface brightness (SB) profiles with \citet[][K66 hereafter]{king66} single-mass models convoluted with a simple model of the ACS Point Spread Function (PSF), as done in \citet{federici07,barmby07}. The central SB was fixed at the value of the innermost point of the observed profile and then we searched for the values of the core radius ($r_c$) and of the concentration parameter ($C={\rm log(r_t/r_c)}$, where ${\rm r_t}$ is the tidal radius, see K66) that minimises $\chi^2$. Second, we fitted straight \citet[][EFF87]{eff87} models to the observed profiles with the curve fit Python library, as done in \citet{gatto21}.
The observed profiles of the clusters, their best fitting curves and the resulting structural parameters are presented and briefly discussed in App.~\ref{app:sbprof}. The two sets of models perform very similarly in the innermost regions of the clusters, while EFF87 appears much better in reproducing the extended outer regions of the clusters. In particular, the external parts of the profiles of C1, C3 and C8 do not seem compatible with tidal truncation, as typical of clusters in dwarf galaxies \citep[EFF87,][and references therein]{gatto21}. It is reassuring to note that integrating the EFF87 profiles we get integrated magnitudes in excellent agreement with those reported in Tab.~\ref{tab:tab1}. On the other hand, core and half-light radii from the fit of K66 model must be considered as our best estimates of the true cluster size  since in this case the effect of the PSF is taken into account. 

Stellar PSF photometry was obtained using the latest version of DOLPHOT \citep{dolphot1,dolphot2}, following the approach  described in \citet{annibali19} and the set-up by \citet{williams14}. Additional details and the adopted selections are described in App.~\ref{app:data}. In the left panel of Fig.~\ref{fig:cmdmap} we show the CMD of the entire ACS sample, that is dominated by a prominent RGB, tipping around F814W$_0\simeq 24.1$ and 
(F606W-F814W)$_0\simeq 1.0$. A significant population of Asymptotic Giant Branch (AGB) stars, tracing intermediate-age populations, is clearly visible above the RGB tip. At colours bluer than the RGB, the sequences of red and blue young core He-burning (He-b) stars can be easily identified, as well as the young Main Sequence (MS), clearly marking the blue edge of the CMD  for 
F814W$_0\la 27.0$. Thanks to the quality of our data, it is easy to select RGB stars on this CMD to obtain a map of the oldest populations that can be identified in our data, displayed in the right panel of Fig.~\ref{fig:cmdmap}. Both the southern and the northern tidal tails are clearly evident in the map, as well as the SW-bending sharp edge of the latter (please, compare with Fig.~\ref{fig:ACSLBC}, right panel). It is extremely interesting and somehow striking, given the distance of the galaxy, that six of the eight clusters listed in Tab.~\ref{tab:tab1} are readily visible in the map as small scale overdensities of RGB stars: the five newly discovered clusters (C1, C2, C3, C5, and C8) and the outermost cluster already found by LEGUS, C7. The CMD of small circular fields around these clusters are shown in Fig.~\ref{fig:cmd} and compared with the CMD of control fields. The comparisons fully support the idea that the clusters are likely old and metal-poor (albeit in some of them the presence of intermediate-age AGB stars cannot be excluded, see App.~\ref{app:cluima}).
It is intriguing to note that C1, C2 and C3 are projected onto the northern tail, somehow following the bending of its sharp edge. The position of C5 is not incompatible with the possible southward prolongation of the Northern tail\footnote{Some of the LEGUS candidate clusters look intriguingly aligned in a north-south narrow sheet southward of C5 and close to C7 and C8, but all of them were classified as low-quality candidates. Indeed, a close inspection of our images confirmed that they actually are background/foreground objects. }. C7 and C8 lie just beyond the south-east edge of the Southern tail. In the following we will focus our attention on these six clusters as they are those most likely associated with the tidal tails\footnote{Although they do not
emerge as obvious overdensities over the strong background of
stars in the main body of the galaxy 
in Fig.~\ref{fig:cmdmap}, also the CMDs of C4 and C6 are dominated by RGB stars and their properties are also typical of classical GCs. The optical spectrum of C6 provided by the Sloan Digital Sky Survey DR17 \citep{ahumada20} suggests that the cluster should have an age in the range 1-4~Gyr.}. In the following we will adopt $D = 4.17 \pm 0.10$~Mpc and E(B-V)=0.01 (see App.~\ref{app:dist}).

It is worth noting that the entire field of view spanned by our ACS images, with a maximum projected distance from the galaxy centre of $\simeq 6.3$~kpc, is dominated by stars of NGC~5238. This means that the tidal tails are immersed into a vaster, low SB halo of which we do not detect the end, where additional star clusters may lie. 

\section{Discussion and conclusion}
\label{sec:discu}

\begin{figure}[ht!]
\center{
\includegraphics[width=\columnwidth]{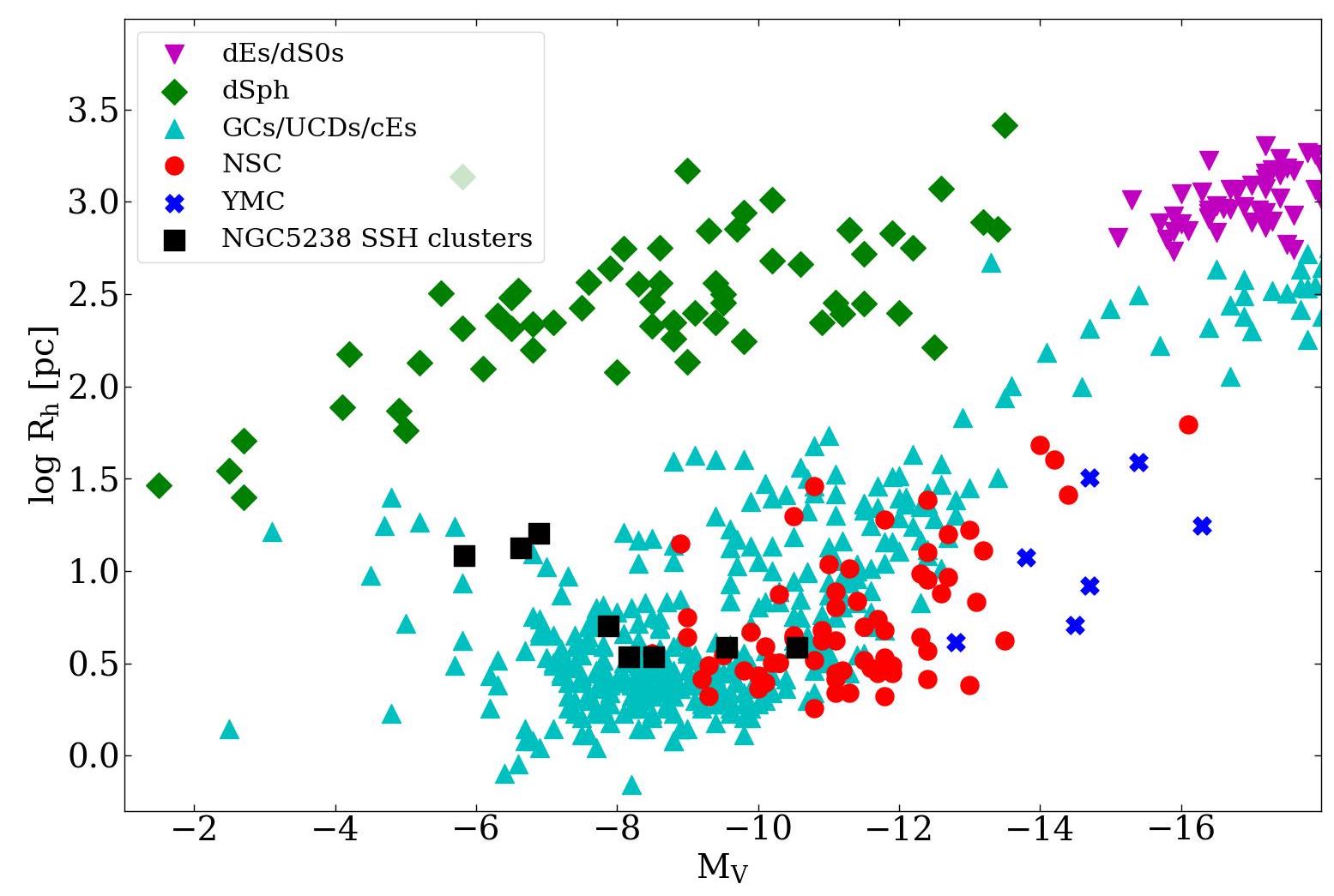}
}
\caption{NGC~5238 clusters listed in Tab.~\ref{tab:tab1} are compared to the sample of pressure-supported stellar systems from \citet{norris14} in the plane opposing the logarithm of R$_{h}$ to the absolute integrated V magnitude. The \citet{norris14} catalogue does not discriminate between GCs, UCDs and compact ellipticals (cEs) but the distribution of the GC/UCD/cE class is dominated by UCDs and cEs for $M_V\le -10.0$ and by GCs for $M_V> -10.0$.}
\label{fig:scale}
\end{figure}

To derive absolute integrated magnitudes in the V band, we transformed the ACS magnitudes into V using Eq.~1 of \citet{galleti06}. The clusters in Tab.\ref{tab:tab1} 
have $-5.8\le M_V\le -10.5$ and half-light radii $3.4~{\rm pc}\le R_h\le 14.5~{\rm pc}$, within the range spanned by classical GCs (see Fig.~\ref{fig:scale}). Adopting the mean mass-to-light ratio of Galactic GCs  \citep[$M/L_V=1.8$,][]{baum20}, the estimated masses lie in the range $4.5\le {\rm log(M/M_{\sun})}\le 6.4$.   We verified that they follow all the GC scaling relations that we were able to check \citep[e.g, C vs. r$_c$, C vs. M$_V$, etc.;][]{djo03}. Their mean colours as well as their CMD  are typical of old and mildly metal-poor globular clusters. 
The fact that six of these bright GCs lie in the extreme outskirts of the galaxy and are correlated (in projection) with the prominent tidal tails discovered by \citet{sacchi24} may suggest that they were associated with the destroyed satellite that produced the tails. 
We have clear 
examples of GCs lying within tidal tails, and therefore being accreted into a larger galaxy along tidal tails, around the MW \citep[e.g.,][and references therin]{mic_sgrclus} and M31 \citep{mackey19} but, to the best of our knowledge, this is the first case in which this is observed to occur in a dwarf galaxy \citep[while accretion of clusters in a dwarf have been suggested by][for NGC~6822]{hwang14}. 

The case of C1 is especially intriguing in this sense. C1, located straight in the middle of the Northern tail, has an absolute integrated V magnitude $M_V=-10.56\pm 0.07$ that is very similar to that of $\omega$~Cen
\citep[$M_V=-10.51\pm 0.05$][]{baum20,baum21}\footnote{\url{https://people.smp.uq.edu.au/HolgerBaumgardt/globular/}}, the brightest GC of the Milky Way, a galaxy with a GC system of more than 150 members and whose stellar mass is $\simeq 500$ times larger than that of NGC~5238 \citep{bh16ARAA}. $\omega$~Cen is generally believed to be the nuclear remnant of a former, and now disrupted, satellite of the MW \citep{ibata19}, in analogy with the case of another very bright GC, M~54 ($M_V=-9.99\pm 0.07$), that is part of the stellar nucleus of the disrupting Sgr~dSph galaxy \citep{mic2008,eugenio10,alfacue2019,alfacue2020}. Indeed, Fig.~\ref{fig:scale} shows that C1 lies in the region of the $M_V$ - log(R$_h$) plane that is dominated by faint NSCs and Ultra Compact Dwarf galaxies \citep[UCD; at least some of which may be the compact central remnant of  tidally stripped galaxies;][N20]{misgeld11,mieske13}. In particular, its luminosity and size are typical of stellar nuclei in dwarf galaxies ($M_{\star}<10^9~M_{\sun}$; see Tab~2 of N20). Also the low ellipticity, $\epsilon=0.07\pm 0.01$\footnote{Average and standard deviation of the measures obtained from the F606W and F814W images with Sextractor.}, is typical of low mass NSCs.
It is tantalising to hypothesise that C1 was in fact the nuclear remnant of the dwarf satellite that was disrupted in the interaction with NGC~5238, producing the tidal tails. 
Indeed this appears as the most likely explanation for the presence of such a massive star cluster so far away from the centre of a dwarf galaxy ($\simeq 2.5$~kpc, in projection, more than five times the galaxy R$_h$\footnote{According to the R$_h$ values reported in the NASA/IPAC Extragalactic Database \url{http://ned.ipac.caltech.edu}}). In this context it is also worth recalling another case of a possible nuclear remnant found into the outskirts of a dwarf galaxy, CL77 in NGC~4449, reported by \citet{annibali12}, as well as the examples of UCDs caught in the process of formation by tidal shredding around spiral galaxies 
reported by \citet{jennings15} and \citet{paudel23}. 
The case of NGC3628-UCD1 is particularly interesting, as the system is embedded in a tidal tail and it is remarkably similar to C1 in size and luminosity. NGC~5238 hosts another star cluster of comparable brightness as C1 close to its optical centre, LEGUS~242 with $M_V=-10.7\pm 0.3$ (from LEGUS photometry). However, this cluster has a much bluer colour than C1, $(F606W-F814W)_0=0.43\pm 0.12$ vs. $(F606W-F814W)_0=0.73\pm 0.03$ and appear to be associated with very strong nebular emission \citep[][]{cannon16}. LEGUS reports a very young age for this cluster ($10^6-10^7$ yr, depending on the version of the catalogue) and, consequently a relatively low mass $M\la 5\times 10^4~M_{\sun}$.

At this stage, only speculations based on uncertain assumptions can be made on the nature of the hypothesised progenitor of C1. According to the relation between the stellar masses of NSCs and the host galaxies (Eq.~1 of N20), if C1 is a stripped NSC the most likely mass of its parent galaxy is $M_{\star}\simeq 10^{8.7}~M_{\sun}$ but, given the large scatter of the relation, progenitor masses down to $M_{\star}\simeq 10^{7.0}-10^{8.0}~M_{\sun}$ are also possible \citep[see also][]{lambert24}, that would imply a (stellar) mass ratio in the range $\sim 0.1-1$ for the merging event, assuming, for simplicity, that the two interacting galaxies have the same M/L ratio. 
Assuming that the six clusters in the tidal tails belonged to the disrupted satellite, its GC specific frequency \citep[$S_N$, see][]{bs06}, will fall within the range observed in nucleated dwarf with $M_V\ga -14.5$ for stellar mass ratios $\ga 0.2$ \citep{millerlotz2007}. A lower mass ratio for the merging is possible, e.g., 
if C1 is itself a low-mass UCD \citep[][]{saifollahi21}, in which case it would be difficult to explain the presence of the other clusters in the tails.  

The tentative model of NGC5238 and its tails by \citet[][]{pascale24} was only constrained by the apparent surface density distribution and by the \ion{H}{i} kinematics. The discovery of these bright clusters in the tidal tails opens the path for obtaining new and more powerful constraints, e.g., by using the clusters to probe the stellar velocity field in the tails, thus opening a powerful window on  the rare and fascinating case of nearby merging between dwarf galaxies.

\begin{acknowledgements}

We are grateful to Angela Adamo and David Cook for their help and assistance in the use of LEGUS catalogues.

MB and  FA acknowledge financial support to this project by INAF, through the PRIN-2023 grant Ob. Fu 1.05.23.05.09 "Dwarf galaxies as probes of the Lambda Cold Dark Matter hierarchical paradigm at the smallest scales" (P.I.: F. Annibali). 

M. Correnti acknowledge financial support from the ASI-INAF agreement n.2022-14-HH-0.
M. Cignoni  acknowledges the support of INFN ”iniziativa specifica TAsP”.
MG acknowledges the INAF AstroFIt grant Fu Ob. 1.05.11.
RP acknowledge the financial support to this research by the Italian Research Center on High Performance Computing Big Data and Quantum
Computing (ICSC), project funded by European Union - NextGenerationEU - and National Recovery and Resilience Plan (NRRP) - Mission 4 Component 2 within the activities of Spoke 3 (Astrophysics and Cosmos Observations).

These data are associated with the HST GO program 17140
(PI: F. Annibali). Support for program number 17140 was provided by NASA through a grant from the Space Telescope Science Institute, which is operated by the Association of Universities for Research in Astronomy under NASA contract 

This work is based on LBT data. The LBT is an international collaboration among institutions in the United States, Italy, and Germany. LBT Corporation partners are the University of Arizona on behalf of the Arizona Board of Regents; Istituto Nazionale di Astrofisica, Italy; LBT Beteiligungsgesellschaft, Germany, representing the Max Planck Society, the Leibniz Institute for Astrophysics Potsdam, and Heidelberg University; the Ohio State University, and the Research Corporation, on behalf of the University of Notre Dame, University of Minnesota, and University of Virginia. We acknowledge the support from the LBT-Italian Coordination Facility for the execution of observations, data distribution, and reduction. 

This research has made use of the SIMBAD database, operated at CDS, Strasbourg, France. This research has made use of the NASA/IPAC Extragalactic Database (NED), which is operated by the Jet Propulsion Laboratory, California Institute of Technology, under contract with the National Aeronautics and Space Administration.

In this analysis we made use of TOPCAT (\url{http://www.starlink.ac.uk/topcat/}, \citealt{Taylor2005}), APT (\url{https://www.aperturephotometry.org} \citealt{apt1,apt2}), DOLPHOT (\url{http://americano.dolphinsim.com/dolphot/},\citealt{dolphot1, dolphot2}).

\end{acknowledgements}


\bibliographystyle{aa} 
\bibliography{refs} 


\begin{appendix}

\section{Properties and isolation of NGC~5238}
\label{app:N5238}

In table~\ref{tab:tab3} we summarise the main properties of NGC~5238, including the tidal indices defined by \citet{kara13} to quantify the possible impact of tides from nearby galaxies \citep[see][for a discussion related to the selection of SSH targets]{annibali16}. Negative values of these indices imply that the effect of tidal forces from known surrounding galaxies is negligible. $\theta_1$ is intended to measure the tidal force exerted on the galaxy by the main disturber while $\theta_5$ combines the effects of the five strongest disturbers. It is interesting to note that both parameters are negative, in the case of NGC~5238.

NGC~5238 is a member of the Canes Venatici I Cloud \citep[CVn~I Cloud;][]{makarov13} that includes several galaxies with distance and line of sight velocity very similar to it. However, according to \citet{kara13} the main disturber of NGC~5238 is the Seyfert galaxy NGC~4736 (M~94) that lies $\simeq 13.0\degr$ apart in projection, corresponding to $\simeq 960$~kpc.
The closest dwarf galaxy is UGC~8331, $\simeq 5.1\degr$ apart in projection, corresponding to $\simeq 370$~kpc. Assuming that this galaxy is exactly at the same line of sight distance as NGC~5238 and a transverse velocity two times the velocity dispersion of the CVn~I Cloud \citep[$\sigma=51$~km/s;][]{makarov13}, it would require $\simeq 3.5$~Gyr to bring the two galaxies at the same position. In summary, it seems highly unlikely that the existing galaxies around NGC~5238 can be responsible for its disturbed morphology; the galaxy appears to be remarkably isolated.

\begin{table}[!htbp]
\centering
\caption{\label{tab:tab3} Main properties of NGC5238.}
{
    \begin{tabular}{lll}
Parameter  & Value & Notes \\
\hline 
$D$      & $4.2\pm 0.1$~Mpc & this work \\
$M_V$      & $-14.3\pm 0.1$ & HyperLeda$^a$  \\
$R_h$      & 236$\pm$8~pc  & (major axis) NED$^b$     \\
$M_{\star}$  & $8.9\times10^7~M_{\sun}$  & \citet{cannon16}  \\
M$_{\ion{H}{I}}$  & $2.2\pm 0.3 \times10^7~M_{\sun}$  & \citet{cannon16}  \\
$M_{\rm dyn}$  & $3\times10^8~M_{\sun}$  & \citet{cannon16}  \\
12+log(O/H)  & 7.96$\pm$0.20  & \citet{mousta06}$^c$  \\
$V_h$  & $232\pm 1~{\rm km~s^{-1}}$   &  \citet{cannon16}   \\
$\theta_1$ & -0.4                     & \citet{kara13} \\
$\theta_5$ & -0.2                     & \citet{kara13} \\
\hline
    \end{tabular}
}
\tablefoot{$^a$ \url{http://atlas.obs-hp.fr/hyperleda/}, from SDSS g and r magnitudes by \citet{aihara11}
transformed as in App.~\ref{app:sbprof}. $^b$ Nasa Extragalactic 
Database \url{http://ned.ipac.caltech.edu}, from SDSS DR5 measures. 
$^c$ See also \citet{marble10} and \citet{cannon16}.
$V_h$ is the heliocentric 
line of sight velocity, measured with \ion{H}{i} observations.
}
\end{table}

\FloatBarrier

\section{Data reduction}
\label{app:data}

We downloaded from the HST archive\footnote{\url{https://mast.stsci.edu/portal/Mashup/Clients/Mast/Portal.html}} the {\it flc} science images, which correspond to the bias-corrected, dark-subtracted, flat-fielded, CTE corrected, GAIA-aligned images. We combined them in a single-stacked, distortion-corrected image (to be used as a reference frame in the photometric data reduction). Then, we used the latest version of DOLPHOT \citep[][]{dolphot1,dolphot2}  to obtain simultaneous multi-filter PSF photometry. We set DOLPHOT parameters using a combination between the default values and those derived by \citep{williams14}. 

To exclude artifacts and spurious detections from the DOLPHOT output, we adopted a series of selection cuts using diagnostic parameters included in the photometric catalog \citep[following, e.g., ][]{annibali19}. In particular, after a first selection where we retained only the sources with  {\tt Object type $\leq$ 1}, photometry quality {\tt flag $\leq$ 2}, and {\tt S/N $>$ 3}, we adopted cuts in the  {\tt sharp} parameter, tracing the difference in extension of the image of a source with respect to a point source, and in the {\tt crowd} parameter, indicative of the degree of contamination by nearby stars, that allows to remove a large fraction of contaminants \citep[see, e.g.][and references therein]{vv124}. The {\tt sharp} cuts have been derived from the {\tt sharp} vs magnitude distribution, selecting sources with {\tt |sharp|} $<$ 0.075 or, after calculating the mean and sigma of the {\tt sharp} distribution in 0.5 magnitudes bins, within $\pm 2.0 \sigma$ from the local mean.
For the {\tt crowd} parameter, we selected sources with {\tt crowd} $<$ 0.1 or within 2.0 $\sigma$ from the local mean, calculated from the mean and sigma of the {\tt crowd} distribution in 0.5 magnitudes bins.

Additional details on the data reduction will be provided in a forthcoming paper where we will discuss in detail the structure and star formation history of the galaxy. 

\newpage
\section{Cluster images and CMDs}
\label{app:cluima}

\begin{figure}[ht!]
\center{
\includegraphics[width=\columnwidth]{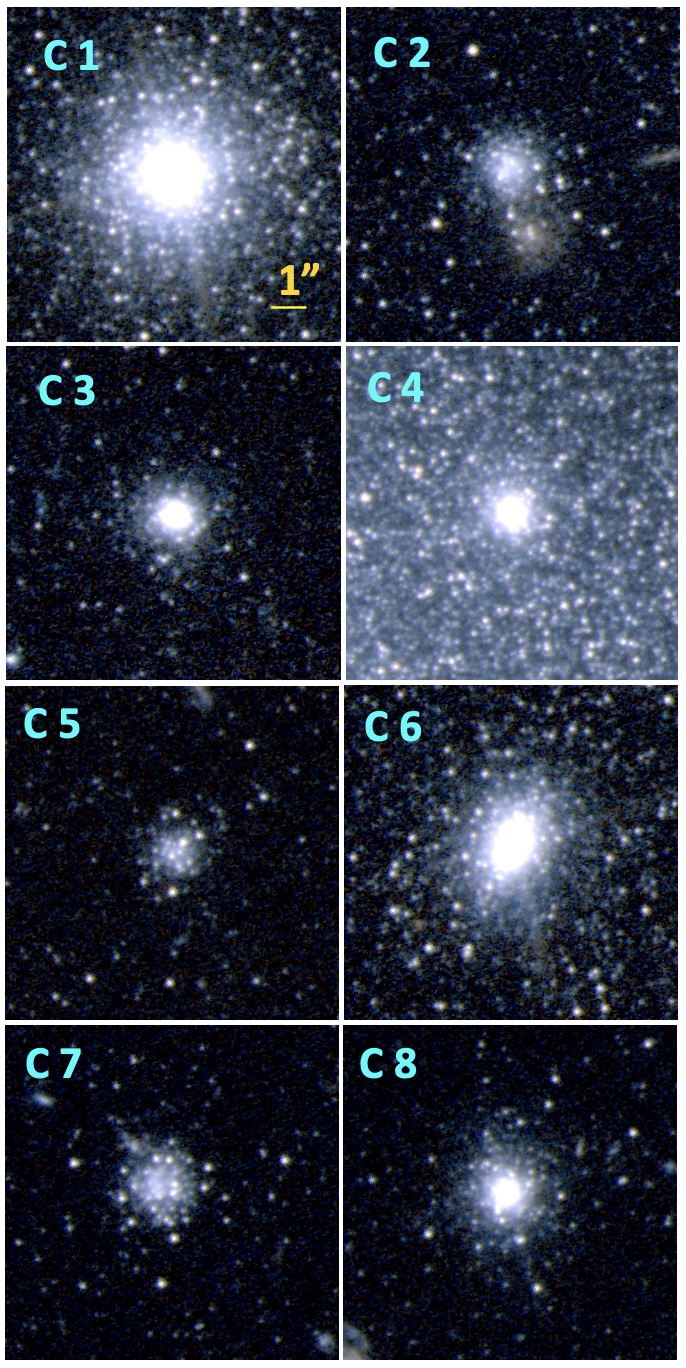}
}
\caption{Stamp images zoomed on a $10\arcsec \times 10\arcsec$ box centred on the cluster for all the eight clusters listed in Tab.~\ref{tab:tab1}. North is up and East to the left. The RGB images have been obtained using the F814W image for the Red channel, the F606W image for the Blue channel, and the sum of the F814W and F606w images for the Green channel.}
\label{fig:stamp}
\end{figure}

\begin{figure}[ht!]
\center{
\includegraphics[width=\columnwidth]{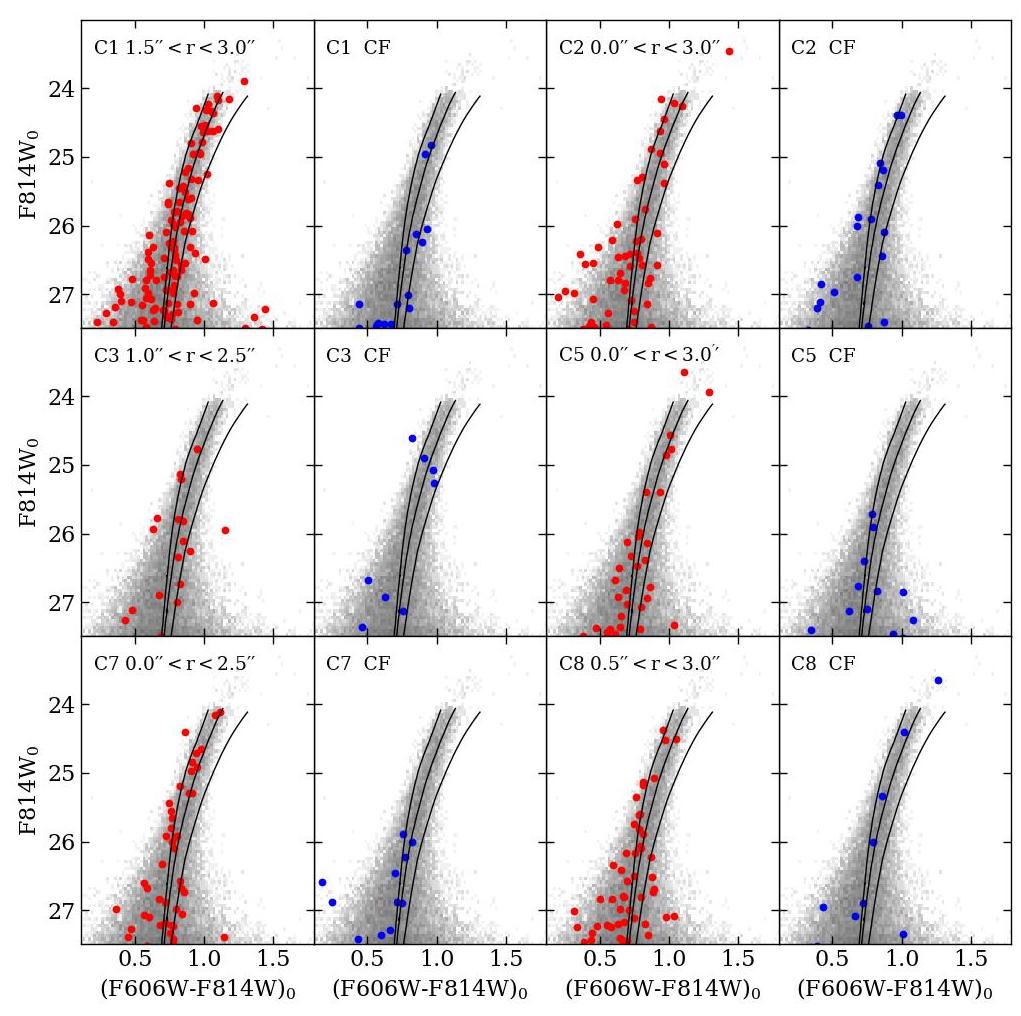}
}
\caption{Colour magnitude diagrams for the six star clusters possibly associated with the tidal tails. In each panel the CMD of the field population in the outskirts of NGC~5238 (r$>70.0\arcsec$) is plotted in grey and the RGB of three Padova isochrones \citep{bressan2012} of age 12.0~Gyr and [M/H]=-2.2, -1.5, -1.0 (from left to right) are plotted as black continuous curves, for reference. For each cluster, stars within a small radial annulus around the cluster centre are plotted as red circles in the "on cluster" CMD, while those of a nearby Control Field (CF) of the same area are plotted as blue circles in the flanking CMD. The limits of the radial annuli are reported in the "on cluster" panels.}
\label{fig:cmd}
\end{figure} 

The RGB colour stamp images zoomed on the eight clusters listed in Tab.~\ref{tab:tab1} are presented in Fig.~\ref{fig:stamp}. There are several features that may be worth noting here: (a) the great extension of C1, with many well resolved stars in the outer regions, (b) the fact that all the clusters appear as partially resolved,
(c) the background galaxy very close to C2, to the south-south-west of its centre, (d) the fact that C2, C5 and C7 are significantly less compact that the other five clusters, with bright stars resolved also very close to the centre, (e) the remarkable elongation of C6 (see App.~\ref{app:sbprof}). All these clusters appear as genuine GCs.

In Fig.~\ref{fig:cmd}, for each of the six clusters possibly associated with the tidal tails, we show the CMD of the stars enclosed within the reported radial annulus centred on the cluster centre (left panels, red filled circles), flanked by the CMD of the stars located within an annulus of the same area but centred $\simeq 20\arcsec$ apart from the cluster (Control Field; CF), carefully avoiding possibly problematic regions (e.g. including bright foreground stars and/or bright or extended background galaxies). Both the stars in the cluster and in the CF region are superimposed to the CMD of the stars in the outskirts of the galaxy (r$>70.0\arcsec$, grey dots), for reference. It may be worth noting that these regions of the galaxy are populated only by old (RGB) and intermediate-age (bright AGB) stars. We have also superimposed to each CMD the RGB of three theoretical isochrones from the Padova set \citet{bressan2012}\footnote{\url{http://stev.oapd.inaf.it/cgi-bin/cmd}} of age 12.0~Gyr and [M/H]=-2.2, -1.5, -1.0, properly shifted to the distance of NGC~5238. From all the comparisons it emerges that the considered clusters are indeed over-densities of RGB stars with respect to the surrounding field, as shown already in Fig.~\ref{fig:cmdmap}, albeit in the case of the most compact cluster (C3) the statistical significance may be low due to the low number of individual stars that can be resolved in the cluster outskirts. In all cases the cluster stars are roughly enclosed between the [M/H]=-1.5 and the [M/H]=-2.2 isochrones, suggesting that the clusters, if old, are quite metal-poor. In the cluster regions of C1, C2 and C3 there are also one, one and two AGB stars brighter that the RGB tip, respectively, that have no counterpart in the CFs. This may suggest that these clusters may in fact have an intermediate age. It is hard to state the statistical significance of this difference given the extremely low numbers and the fact that field AGB stars are distributed all over the galaxy.

\section{Surface brightness profiles}
\label{app:sbprof}

\begin{figure}[ht!]
\center{
\includegraphics[width=\columnwidth]{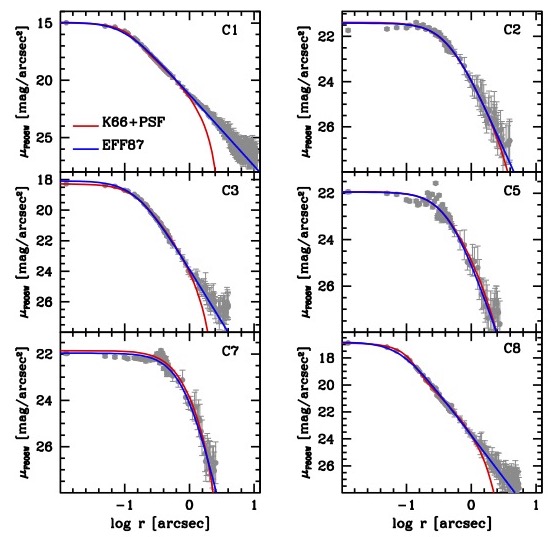}
}
\caption{Surface brightness profiles of the clusters possibly associated with the NGC~5238 tidal tails with over-plotted the best-fit K66 model convolved with a simple model of the HST PSF (red) and EFF87 (blue) models.}
\label{fig:prof}
\end{figure} 

The observed SB profile together with the best fit K66 and EFF87 models for the six clusters possibly associated with the tidal tails of NGC~5238 are displayed in Fig.~\ref{fig:prof}.  The adopted EFF87 profiles have the form:

\begin{equation}
{\rm \mu(r) = \mu_0 +1.25\gamma  log} \left( {\rm 1+\frac{r^2}{\alpha^2}} \right)
\end{equation}

where $\mu_0\equiv \mu(0)$ \citep{gatto21}.

As done in \citet{gatto21}, the fit of the EFF87 is limited to the SB points out to a limiting radius $r_f$ that approximately coincide to the last points of the observed profile shown in each panel of Fig.~\ref{fig:prof}.
The structural parameters of the studied clusters, derived from the SB profile fits shown in Fig.~\ref{app:sbprof} are reported in Table~\ref{tab:tab2}. 
The half-light radii have been obtained by direct integration of the best-fit profiles.

As a sanity check on the remarkably bright integrated magnitudes of C1, we used APT to measure them also in the SSH LBC images. We obtain g$=17.72\pm 0.06$ and r$=17.22\pm 0.05$, in the Sloan Digital Sky Survey ABMAG photometric system \citep{fukugita96}, calibrated as described in \citet{ssh_pap1}. With the same assumptions on distance and reddening as above, the corresponding absolute magnitudes are $M_g=-10.42 \pm 0.08$ and $M_r=-10.90 \pm 0.07$. The colour $(g-r)_0=0.48\pm 0.08$ is typical of classical MW GCs \citep{peacock2010}. Using the transformation

\begin{equation}
    V = g - 0.5784(g-r) -0.0038 
\end{equation}

by Lupton (2005)\footnote{\url{https://www.sdss3.org/dr9/algorithms/sdssUBVRITransform.php\#Lupton2005}} we obtain $M_V=-10.7\pm 0.1$, compatible with the measure obtained from HST data and, above all, providing a fully independent confirmation of the high luminosity of C1.

This letter is focused on the six clusters that appear to correlate more strictly to the tidal tails of NGC~5238. C4 and C6 were included in our list because they appeared evident in the visual inspection of the galaxy outskirts that lead to the discovery of the new clusters. However, in the present analysis, they play the precious role of a bridge between us and LEGUS since they allowed us to verify that, for these clusters, our newly derived integrated magnitudes are in good agreement with the independent measures by LEGUS (see Sect.~\ref{sec:analysis}). In particular, the differences in the integrated magnitudes between our measures and those by LEGUS are $\Delta F606W=0.01\pm 0.10$ and $\Delta F814W=-0.06\pm 0.10$ for C4, and $\Delta F606W=-0.14\pm 0.09$ and $\Delta F814W=-0.22\pm 0.09$ for C6. The fact that our magnitudes are slightly brighter than those by LEGUS is due to the fact that we adopted large apertures, purposely suited to include the contribution by resolved stars, while Sextractor, by definition, should have considered only the unresolved body of the clusters when computing the magnitudes within the LEGUS pipeline. Indeed, the magnitudes we derived from our images with Sextractor more closely match the LEGUS values than those obtained with APT.

Finally, we note that also C6 is quite a bright cluster ($M_V=-9.55\pm 0.06$), with properties overlapping with those of the faintest NSCs, and it also displays a significant ellipticity, $\epsilon\simeq 0.25$, as measured with Sextractor. 

\begin{table*}[!htbp]
\caption{\label{tab:tab2} Structural parameters of the analysed clusters}
{
    \begin{tabular}{cccccccccc}

  SSH           &  r$_c^{K66}$  &   R$_h^{K66}$  &   C$^{K66}$   & $\mu_0^{K66}$ &   $\mu_0^{EFF87}$	  &    $\alpha$	    &	$\gamma$ & R$_h^{EFF87}$ & log~L$_{F606W}^{EFF87}$	  \\    
    id          & arcsec& arcsec && mag/arcsec$^2$ & mag/arcsec$^2$ &   arcsec      &               &    arcsec  & L$_{\sun}$    \\
\hline 
  C1            & 0.06$\pm$0.01  &  0.19$\pm$0.03  &  1.70 & 14.97 & 14.93$\pm$0.12 & 0.098$\pm$0.005 & 2.50$\pm$0.01 &  0.30$\pm$0.19 & 6.22$\pm$0.07\\
  C2            & 0.37$\pm$0.06  &  0.79$\pm$0.12  &  1.35 & 21.37 & 21.41$\pm$0.06 & 0.440$\pm$0.024 & 2.59$\pm$0.07 &  0.74$\pm$0.10 & 4.86$\pm$0.07\\
  C3            & 0.12$\pm$0.01  &  0.25$\pm$0.02  &  1.35 & 18.28 & 18.08$\pm$0.08 & 0.147$\pm$0.006 & 2.76$\pm$0.03 &  0.28$\pm$0.06 & 5.13$\pm$0.05\\
  C4 (11)	& 0.06$\pm$0.01  &  0.17$\pm$0.03  &  1.60 & 16.74 & 16.45$\pm$0.09 & 0.077$\pm$0.003 & 2.45$\pm$0.02 &  0.21$\pm$0.06 & 5.45$\pm$0.06\\ 
  C5            & 0.34$\pm$0.08  &  0.60$\pm$0.14  &  1.20 & 21.94 & 21.95$\pm$0.11 & 0.440$\pm$0.035 & 3.20$\pm$0.13 &  0.57$\pm$0.14 & 4.34$\pm$0.09\\
  C6 (591)	& 0.06$\pm$0.01  &  0.19$\pm$0.03  &  1.70 & 15.83 & 15.75$\pm$0.10 & 0.100$\pm$0.004 & 3.60$\pm$0.01 &  0.25$\pm$0.08 & 5.83$\pm$0.06\\
  C7 (617)  	& 0.69$\pm$0.16  &  0.66$\pm$0.15  &  0.70 & 22.03 & 21.96$\pm$0.06 & 0.899$\pm$0.065 & 4.94$\pm$0.33 &  0.67$\pm$0.09 & 4.56$\pm$0.08\\
  C8            & 0.04$\pm$0.01  &  0.17$\pm$0.04  &  1.90 & 16.89 & 16.82$\pm$0.15 & 0.081$\pm$0.005 & 2.53$\pm$0.02 &  0.23$\pm$0.17 & 5.27$\pm$0.08\\
\hline
    \end{tabular}
}
\tablefoot{Numbers in parentheses in the first column are LEGUS id numbers, when available. Central surface brightness values are not corrected for interstellar extinction. 
log~L$_{F606W}^{EFF87}$ has been derived by integration to infinity of the best-fit EFF87 model and assuming $M_{F606W,\sun}=4.66$ from
\url{https://mips.as.arizona.edu/~cnaw/sun.html}.
The K66 parameters have 
been derived by fitting K66 models convolved with a simple model of the HST PSF to the observed profiles, while EFF87 
parameters are derived by fits not taking into account the effect of the PSF on the profile.  The uncertainty in $r_c^{K66}$ has been computed as the range over which the $\chi^2$ value increases by a factor of 2 with respect to the minimum. C6 is significantly flattened ($\epsilon\simeq 0.25$, as measured with Sextractor) hence the profile we derived with circular apertures may not be an adequate representation of the real profile.
}
\end{table*}
\FloatBarrier

\section{Distance and reddening}
\label{app:dist}

We derived a new distance estimate for NGC~5238 from the RGB Tip (TRGB) using our ACS photometry. Following exactly the same procedure as
\citet{bptip24} we obtain $F814W_0^{TRGB}=24.10\pm 0.01$ and $(F606W-F814W)_0^{TRGB}=1.13\pm 0.06$. Adopting, from the same authors,  $M_{F814W_0}^{TRGB}= -3.996 \pm 0.045$ for the RGB Tip of the Small Magellanic Cloud that have the same colour, we get $(m-M)_0=28.10 \pm 0.05$, corresponding to $D = 4.17 \pm 0.10$~Mpc. At a first glance this is significantly smaller than $D = 4.51 \pm 0.06$~Mpc by \citet{tully09} but, in fact, this is based on a very different calibration and the uncertainty on the calibration is completely neglected, while it is fully taken into account by \citet{bptip24}. Moreover, \citet{tully09} derive $F814W_0^{TRGB}=24.20\pm 0.02$, that is hardly compatible with our CMD, even at a first glance. On the other hand, the LEGUS collaboration reports $D = 4.43 \pm 0.34$~Mpc \citep{legus_cmd,cook23}, compatible with our measure, within the uncertainties. We do not enter here in the discussion of the merit of the various distance estimates, we simply adopt our own one because is smaller than the others, a conservative choice in the derivation of the absolute integrated magnitude of the clusters in the sense that the values that we obtain are unlikely to be inflated by the adoption of an overly large distance.

The \citet{sfd98} reddening maps show that the foreground extinction is very low and uniform over the field covered by our ACS images and in the extreme outskirts of the galaxy, where the clusters we are interested in are located, there should be no contribution from internal extinction. Consequently we adopt $E(B-V)=0.01$, re-calibrating the mean value from the \citet{sfd98} maps according to \citet{schlafly2011}. As a sanity check, these results have been confirmed using the extinction maps by \citet{vergely22}.
We adopt the reddening laws reported in \citet{bptip24}.

\end{appendix}

\end{document}